\documentclass[twocolumn,prl,amsmath,amssymb,showpacs,superscriptaddress,floatfix]{revtex4}
\usepackage{graphicx}
\usepackage{bm}
\usepackage{lineno,hyperref}
\usepackage{epstopdf}
\usepackage{epsf}
\usepackage{xcolor}
\begin{document}
\title{Comparative study of melting of graphite and graphene}

\author{Yu. D. Fomin \footnote{Corresponding author. E-mail: fomin314@mail.ru} and V. V. Brazhkin}
\affiliation{ Institute for High Pressure Physics RAS, 108840
Kaluzhskoe shosse, 14, Troitsk, Moscow, Russia}

\date{\today}


\begin{abstract}
The melting lines of graphite and liquid carbon have been studied for a long time. However, numerous controversies still remain in this field; for instance, different experiments give different melting temperatures.
In this work, we explore the melting lines of graphite and graphene by means of classical and ab initio molecular dynamics. We show that the empirical models fail to reproduce the properties of liquid carbon properly. However, both empirical and ab initio simulations evidence the presence of smooth structural crossover in liquid carbon. We also show that the "melting" of graphene discussed in previous works on computer simulation is indeed sublimation and propose a method to simulate genuine melting of graphene. The true melting temperature of graphene appears to be close to the melting temperature of graphite.
\end{abstract}


\maketitle

\section{1. Introduction}

Melting of graphite and diamond and the nature of liquid carbon are two
long-standing problems of condensed matter physics. However, in
spite of very intensive research, both problems still remain
unsolved. The first attack on the problems appeared as early as
1911 \cite{gr1911}. In 1923, the experimental works by Pirani and Alterthum
was published \cite{pirani}. The first theoretical calculation of the
phase diagram of carbon was made by Leipunsky in 1939
\cite{leipunsky}. Extrapolating the experimental data available at
that time and introducing some assumptions (e.g., that the
volume jump at melting of graphite does not exceed $5 \%$), the
estimated the melting point of graphite to be $T_m=4000$
K at a moderate pressure. Interestingly, Leipunsky estimated the melting point at a pressure as high as 10 GPa to be 4240 K;
i.e., the slope $dT/dP$ of the melting curve is very small.
However, all these attempts both experimental and theoretical were
rather inaccurate.

The most famous experimental evaluation of the melting curve of
graphite was performed by Bundy \cite{bundy}.
According to his results, the melting temperature is $4100$ K at
A pressure of 0.9 GPa, increases to 4600 K at 7 GPa and decreases back to 4100 K at 12.5 GPa. The presence of a maximum
on the melting curve of graphite was confirmed in the works by
Fateeva and Vereschagin \cite{fv} and Togaya\cite{togaya}.
However, although all three works report that the melting curve
demonstrates a maximum, the melting temperatures are very different
in all these works.

Numerous experimental attempts to evaluate the melting curve of
graphite were carried out (see, e.g., Ref. \cite{savv} and references therein). However, the problem mentioned above still remains: different experiments give different melting points. The
estimates of the melting temperature range from approximately
4000 to 5000 K.

Experimental difficulties with the phase diagram of carbon  also stimulated numerous attempts to calculate it in computer simulation. The
simulation of carbon was mostly performed with so-called Bond
Ordered Potentials (BOPs) and ab initio methods. A liquid--liquid phase transition (LLPT) in carbon was predicted in the series of
works by Glosli and Ree (see \cite{llpt} and references therein), where the famous Brenner potential was used. However, the existence of the LLPT was
immediately doubted by other authors who employed other empirical
potentials (for example, LCBOP) and ab initio simulations.
Currently, it is commonly accepted that no LLPT takes
place in carbon and its appearance in the Brenner model is due to
incorrect evaluation of torsional interactions in this model.

Several other empirical potentials were used to calculate the melting line of graphite.
In Ref. \cite{lcbop}, a long-range carbon BOP (LCBOP) was introduced. Further, this model was modified in Ref. \cite{lcbopiplus} (LCBOPI+ model). The phase diagram of this model is given in Ref. \cite{lcboppluspd}.
In this model, the melting line of graphite has a positive slope $dT/dP$ in the whole range of pressure.
The melting temperature of graphite varies from $T=3800$ K at $P=2$ GPa to $T=4250$ K at
$P=16.4$ GPa (the graphite--diamond--liquid triple point). The properties of liquid carbon were
also studied within this model and no LLPT was observed.

This model was further improved in Ref. \cite{lcbopii} (LCBOPII potential). In Ref. \cite{lcbopii-liq}, the properties of liquid carbon were studied within the LCBOPII model and no LLPT was observed. The phase diagram of the LCBOPII model was calculated in Ref. \cite{colonna}. A particularly interesting prediction of the LCBOPII model is that the melting temperature of graphite is almost independent of the pressure and is equal to $T_m= (4250 \pm 50)$ K. The authors also calculate the specific volume of graphite and liquid carbon along the melting line and show that although both volumes of graphite $V_g$ and liquid carbon $V_l$ decrease with increasing pressure, the equality $V_l \approx V_g$
holds along the whole melting line; i.e., $\Delta V \approx 0$, which, according to the  Clausius–-Clapeyron relation, leads to zero slope $dT/dP$.

Another model widely used to investigate carbon and hydrocarbons is so-called AIREBO potential \cite{airebo}. The melting curve of this model was calculated in Ref. \cite{orekhov} by means of the moving interface method. As in the LCBOPII model, the melting temperature of AIREBO graphite almost does not depend on pressure. In the range of pressure from 2 to 12 GPa, the melting temperature is $T_m= (3640 \pm 150)$ K, which is slightly below the value obtained within the LCBOPII model. No LLPT was also observed in this model.

In Ref. \cite{tb}, tight-binding simulation of liquid carbon is presented. This method can be thought as something intermediate between empirical potentials and ab initio methods in the sense of both the simulation burden and the accuracy of calculations. The tight-binding simulation also does not give any evidences of the LLPT in carbon.
The ab initio simulations of liquid carbon also eliminate the possibility of the LLPT. However, no firm ab initio estimate of the melting point of graphite has been reported.

A new approach to the construction of empirical models based on machine learning methods has been proposed recently. This model was implemented for amorphous carbon in Ref. \cite{gap} (GAP potential) The results obtained with this model are in good agreement with ab initio data, but it is possible to simulate larger systems and longer times, which is especially important for investigation of amorphous carbon. Although the GAP model gave very encouraging results for the properties of amorphous carbon, the melting line of this model has not yet been calculated.

One can see that simple empirical models of carbon, such as the Brenner potential, predict the appearance of the LLPT, while more elaborate models, such as tight-binding and ab initio simulations, do not show the LLPT. For this reason, it is a common view now that the LLPT is an artefact of simple models and it should not take place in real carbon.
However, the conclusions on the presence or absence of the LLPT in carbon are usually based on two foundations: equations of state (EoSs) and hybridization of atoms, which is related to the number of nearest neighbors. If the LLPT is present, then EoSs demonstrate the van der Waals loop and the hybridization changes from dominant sp2 at low densities to dominant sp3 at high ones. However, the accuracy of calculations of both EoSs and hybridization in ab initio simulations is doubtful too. For instance, EoSs calculated in several works by ab initio molecular dynamics simulation are compared in Ref. \cite{dozhdikov}. Agreement between different publications is rather poor. Moreover, the best agreement is achieved between the results obtained in \cite{dozhdikov} and \cite{french} although the simulations in the former and latter works are performed for $T=6000$ and $7000$ K, respectively. The situation with hybridization is very similar. The results obtained in Ref. \cite{dozhdikov} at $T=6000$ K are in better agreement with the results of the GAP model at $T=5000$ K \cite{gap} than with other ab initio studies at the same temperature \cite{dozhdikov}. Thus, the accuracy of the DFT results are doubtful because these results obtained with different functionals differ from each other larger than the results obtained for temperatures differing by 1000 K.

Although the modern empirical models of carbon, tight-binding and ab initio simulations do not demonstrate the LLPT in carbon, the origin of almost zero slope of the melting line of graphite still remains unclear. There are just a few other substances with the slope of the melting line below $50$ $K/GPa$ in a wide range of pressure. This phenomenon can be possibly explained in regular solution theory implemented to study maxima on the melting line \cite{rapaport}. However, the LLPT should take place in this case. Another possibility is the existence of smooth structural crossover in the liquid, which can also be called smooth LLPT. Unlike the "real" LLPT, smooth crossover is not accompanied by a density jump, but the structure of the liquid changes smoothly. In this case, some unusual behavior of structural characteristics, for example, the position and height of the first peak of radial distribution functions (RDFs) can appear. Below, the "real" LLPT will be called the sharp LLPT, whereas the crossover will be referred to as the smooth LLPT.

Another actual problem is the melting of graphene. It was simulated within the LCBOPII model in Refs. \cite{zakh,los} and AIREBO in Refs. \cite{orgr}. Two interesting results are reported in these studies. First, upon "melting," graphene forms a phase composed of linear chains. For this reason, both groups that performed LCBOPII and AIREBO simulations claim that this process is some kind of a polymerization process or, probably, the formation of carbine rather than real melting. However, they propose to remain the term melting for simplicity.
The second point of interest is that the "melting" temperature of graphene appears to be higher than that for graphite. An explanation of this property proposed in Refs. \cite{orgr} is as follows: when graphite is heated, some atoms leave a plane and form bonds with atoms of another plane. Formation of such bridges is responsible for decomposition of graphite. In the case of graphene, no bridges can apparently be formed. Note that the "melting" temperature of graphene obtained wihtin the LCBOPII model in Ref. \cite{zakh} is $T=4900$ K, while the same group reported in Ref. \cite{los} the "melting" temperature $T=4510$ K.
These two publications employ different methodologies: a direct observation of the structural changes is taken as the melting point in the former, whereas in the latter, the authors extrapolate the results to lower temperatures using the Classical Nucleation Theory. All these publications \cite{zakh,los,orgr} conclude that graphene demonstrates the highest melting temperature among the elemental substances.

In this work, we examine the behavior of liquid carbon within the empirical AIREBO model and by ab initio simulations. We show that the AIREBO model does not describe liquid carbon properly and, therefore, cannot describe the melting line. We also study the "melting" of graphene at both nearly zero and elevated pressures and show the inconsistency of the results for melting of graphene with those for graphite. Our results show that the collapse of graphene observed in previous simulations is indeed due to sublimation rather than to melting.

\section{2. System and Methods}

In this study, we simulate the properties of liquid carbon by means of classical molecular dynamics (MD) within the AIREBO model and in ab initio MD with the PBE functional.

For classical MD, we use a system of 4096 particles in a cubic box. The temperature is varied from $T_{low}=4000$ K, which is slightly above the melting line reported in Ref. \cite{orekhov}, to $T_{high}=6000$ K. The densities are varied from $\rho_{min}=2$ $g/cm^3$ to $3$ $g/cm^3$ with the step $\Delta \rho=0.05$ $g/cm^3$. In the case of $T=6000$ K, we enlarged the interval of densities up to $4$ $g/cm^3$. The time step is set to 0.1 fs. At each point, the system is simulated at $10^7$ steps. The first half of the simulation is used for equilibration. In the second half, we perform the calculations of averages. We calculate the pressure as a function of density and temperature (EoS) and the internal energy of the system. By differentiating the internal energy along isochores, we also calculate the specific heat at constant volume.

To analyze the structure of the system, we calculate the radial distribution functions (RDFs, $g(r)$. The average number of nearest neighbors is calculated from the RDF: $NN=4 \pi n \int_{0}^{r_{min}} r^2g(r)dr$, where $NN$ is the number of nearest neighbors, $n=N/V$  is the number density of the system and $r_{min}$ is the position of the first minimum of the RDF. The position and the height of the first maximum of the RDF are also monitored.

The classical MD simulations were performed with the LAMMPS simulation package \cite{lammps}.

For the ab initio simulation, we use the Quantum Espresso package \cite{qe} and the PBE functional. The energy cutoff for wave functions is set to 40 Ry. The cutoff for the charge density is 160 Ry. The system consists of 54 atoms in a cubic cell. We are primarily interested in the RDFs of the system, in particular, the first maximum and minimum of $g(r)$. For this reason, we do not intend to use very large systems,
because the RDF is not sensitive to the system size. Only the $\Gamma$ point is used for calculations in the $k$-space. The time step is set to 20 au ($0.986$ fs). Equilibration takes 10000 steps and calculations of averages take more than 2000 steps. The temperature and density points used in Quantum Espresso (QE) calculations are the same that in classical MD ones. The initial configuration was extracted from that of the AIREBO model at
$T=6000$ K and given density. After that, the system relaxed in the ab initio simulation for more than 2000 steps at the same temperature to
completely eliminate the effect of the initial structure. The final structures of these simulations were used as initial ones for all ab initio simulations.

\section{3. Results and Discussion}

We start the discussion with the results for the EoS of the system. The
melting temperature of carbon modeled by the AIREBO potential is
$T_m=3640 \pm 150$ K \cite{orekhov}. Although we simulate the system from $T_{min}=4000$ K to $T_{max}=6000$ K, most of the results are reported for $T=6000$ K, which is inside the liquid region of the phase diagram for all models and experimental studies. At the same time, this temperature is below the liquid—liquid critical point obtained in \cite{llpt}; therefore, it is low enough to expect structural changes in the liquid phase, if they exist.

\begin{figure}
\includegraphics[width=6cm,height=6cm]{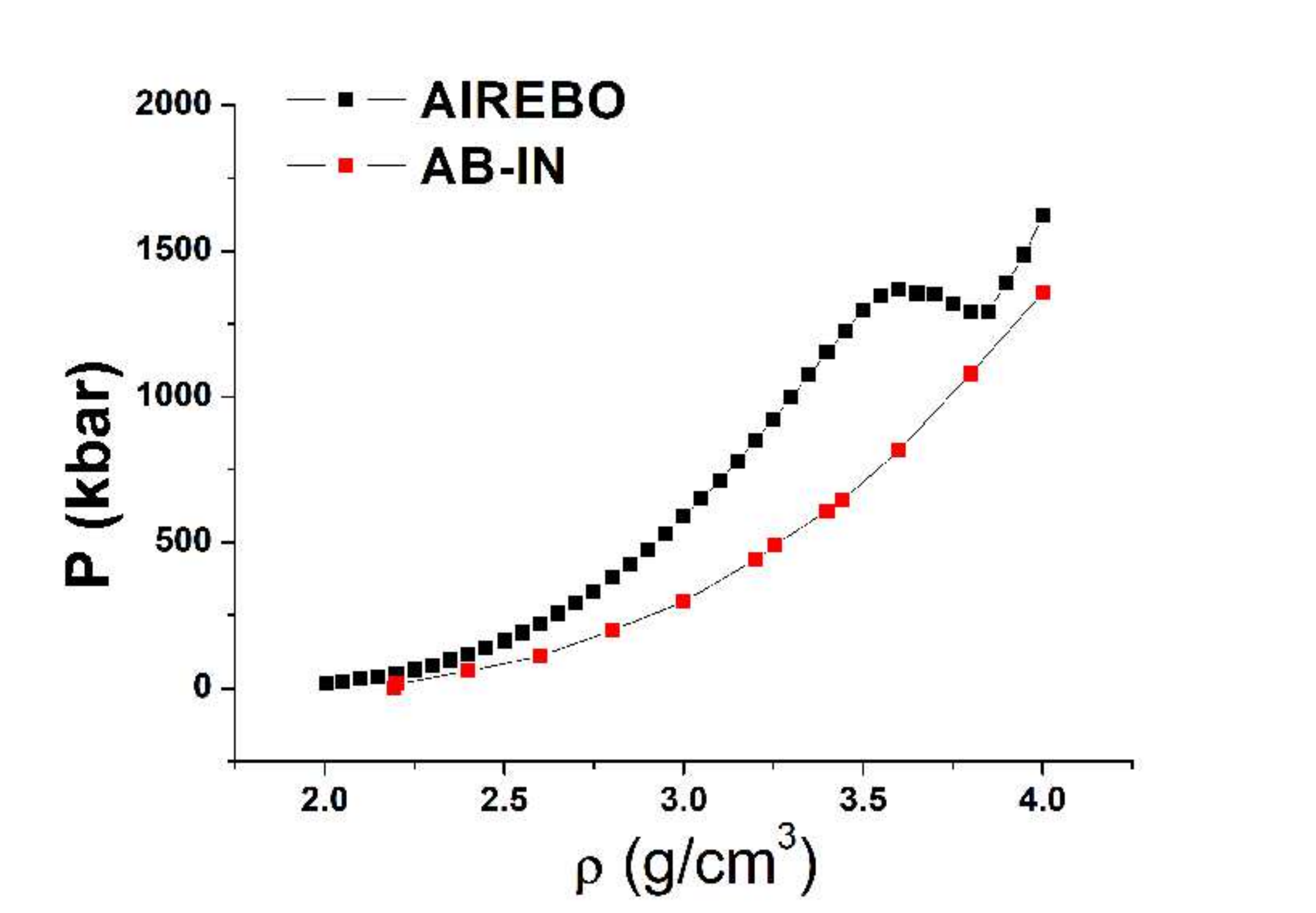}%

\caption{\label{eost6000}Equation of state of liquid carbon at $T=6000$ K obtained within the AIREBO model and in ab initio calculations.}
\end{figure}

Figure \ref{eost6000} shows the equations of state of the AIREBO model and from ab initio simulations at $T=6000$ K up to the density $\rho=4g/cm^3$. Surprisingly, we are not aware of any work where the EoS of the AIREBO model was calculated up to such high densities. One can see that the EoS demonstrates a van der Waals loop, which means that a first order LLPT takes place in the system. However, this LLPT is different from that obtained in Ref. \cite{llpt} because it appears at strongly different densities. No peculiarities are observed in the ab initio EoS. This is similar to the situation with the LLPT predicted in the Brenner model: the LLPT was observed in the empirical model, but not in ab initio simulations. We believe that this phase transition is also a property of the AIREBO model, but not of real carbon. Below, we justify this assertion.

\begin{figure}
\includegraphics[width=6cm,height=6cm]{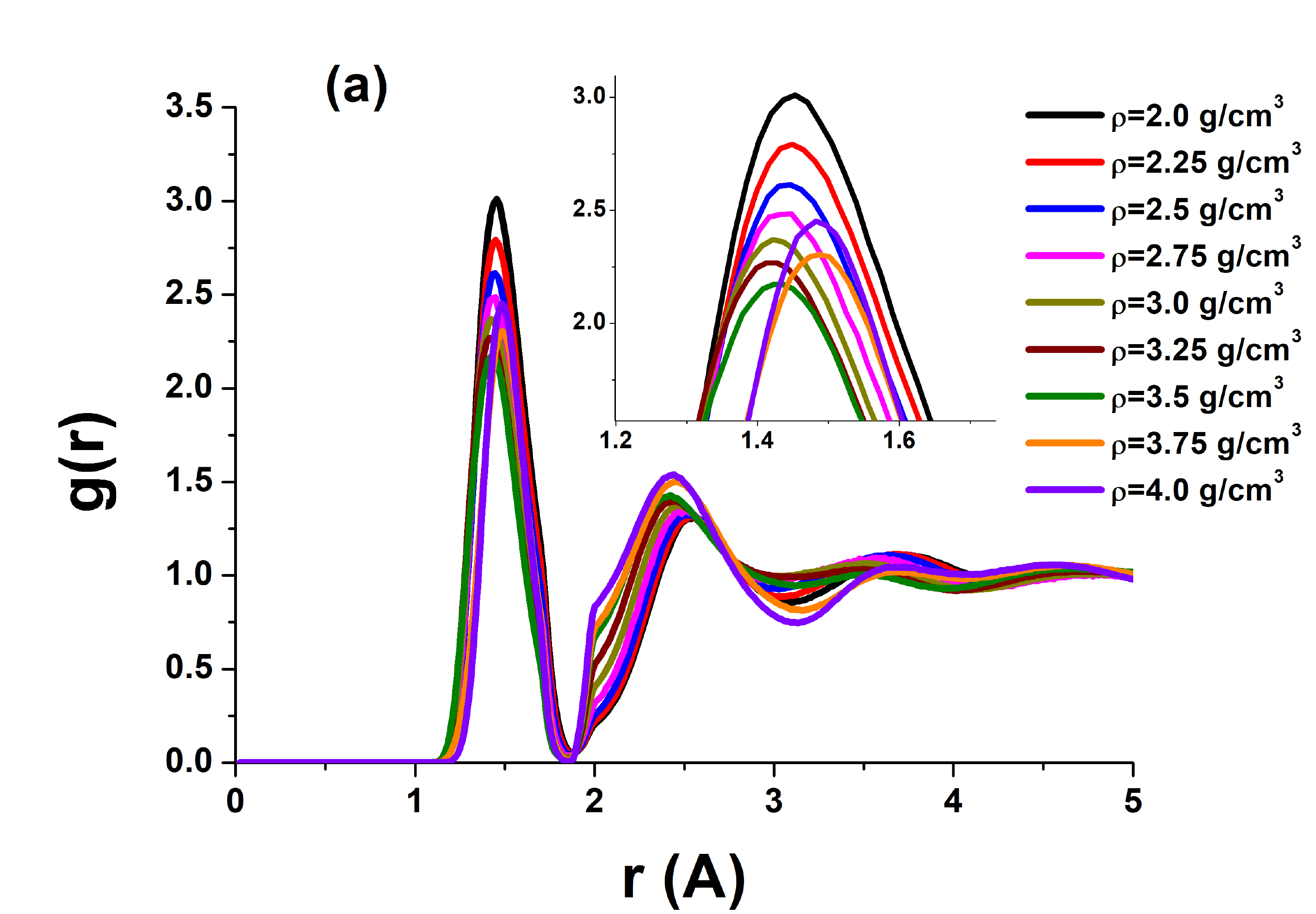}%

\includegraphics[width=6cm,height=6cm]{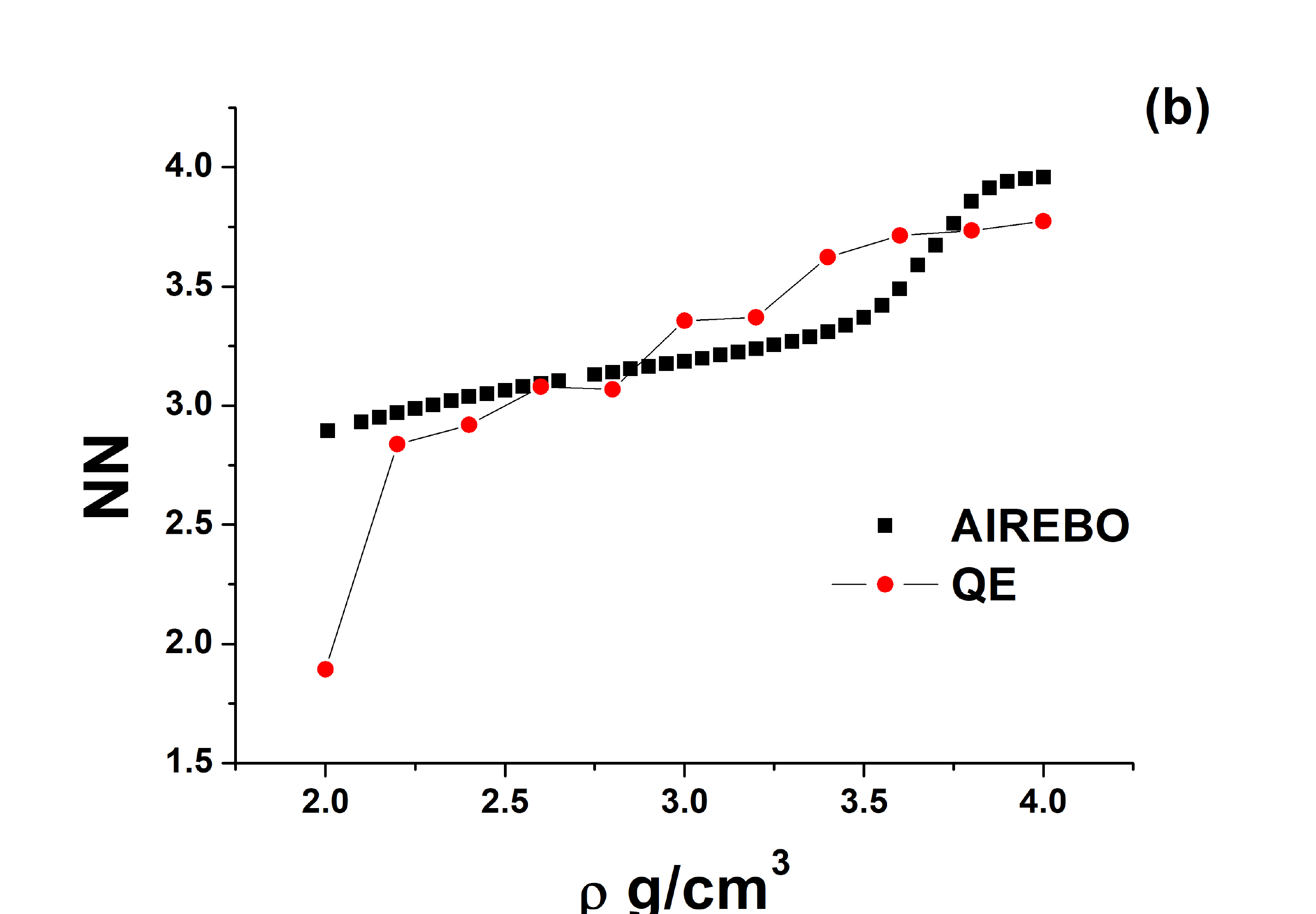}%

\caption{\label{rdft6000}(a) Radial distribution functions of AIREBO model at
$T=6000$ K. (b) Number of nearest neighbors $NN$ in the same system in comparison with the value obtained in ab initio simulations (QE).}
\end{figure}

Usually, when a sharp LLPT in liquid carbon is discussed, one seeks a change in hybridization of the atoms. The hybridization is determined
from the number of nearest neighbors $NN$. Four, three and two nearest neighbors correspond to sp3, sp2 and sp hybridization, respectively. The number of nearest neighbors $NN$ can be determined from the RDF of the system. Figure \ref{rdft6000} shows (a) the RDFs of AIREBO carbon at
$T=6000$ K and (b) $NN$. One can see that $NN$ increases smoothly
with the density, but the increase becomes very rapid in the region of the loop of the EoS and $NN$ at $\rho=4g/cm^3$ becomes equal to 4.

The behavior of $NN$ obtained in the ab initio simulation is different. It is equal to 2 at $\rho=2.0$ $g/cm^3$ and jumps to $2.83$at at $\rho=2.2$ $g/cm^3$. At a further increase in the density, $NN$ increases smoothly to $3.77$. No sharp transition is observed.

However, the RDFs of the system demonstrate more unusual features. In a simple liquid, the position of the first peak of the RDF is lower at a higher density. At the same time, the height of the first peak increases with the density (see, e.g., Ref. \cite{neon} for the experimental and simulation study of the position and height of the first peak of the RDF in neon). However, the inset of Fig. \ref{rdft6000} (a) demonstrates that the situation is more complex in the case of the AIREBO model of carbon. Figure \ref{g1r1} shows the height and position of the first peak in the AIREBO model of carbon. One can see that the height of the first peak $g_1$ decreases anomalously with increasing density up to $\rho=3.6g/cm^3$, i.e., until the LLPT in the system. The behavior of the position of the first peak of the RDF $r_1$ is more simple. It decreases with increasing density up to $\rho=3.2g/cm^3$, then remains nearly constant up to $\rho=3.4g/cm^3$ and increases rapidly at the LLPT up to the density $\rho=3.85g/cm^3$. Then, it again decreases with increasing
density. It means that the behavior of $r_1$, except for the region of the LLPT, coincides with that in a simple liquid.

\begin{figure}\label{g1r1}
\includegraphics[width=6cm,height=6cm]{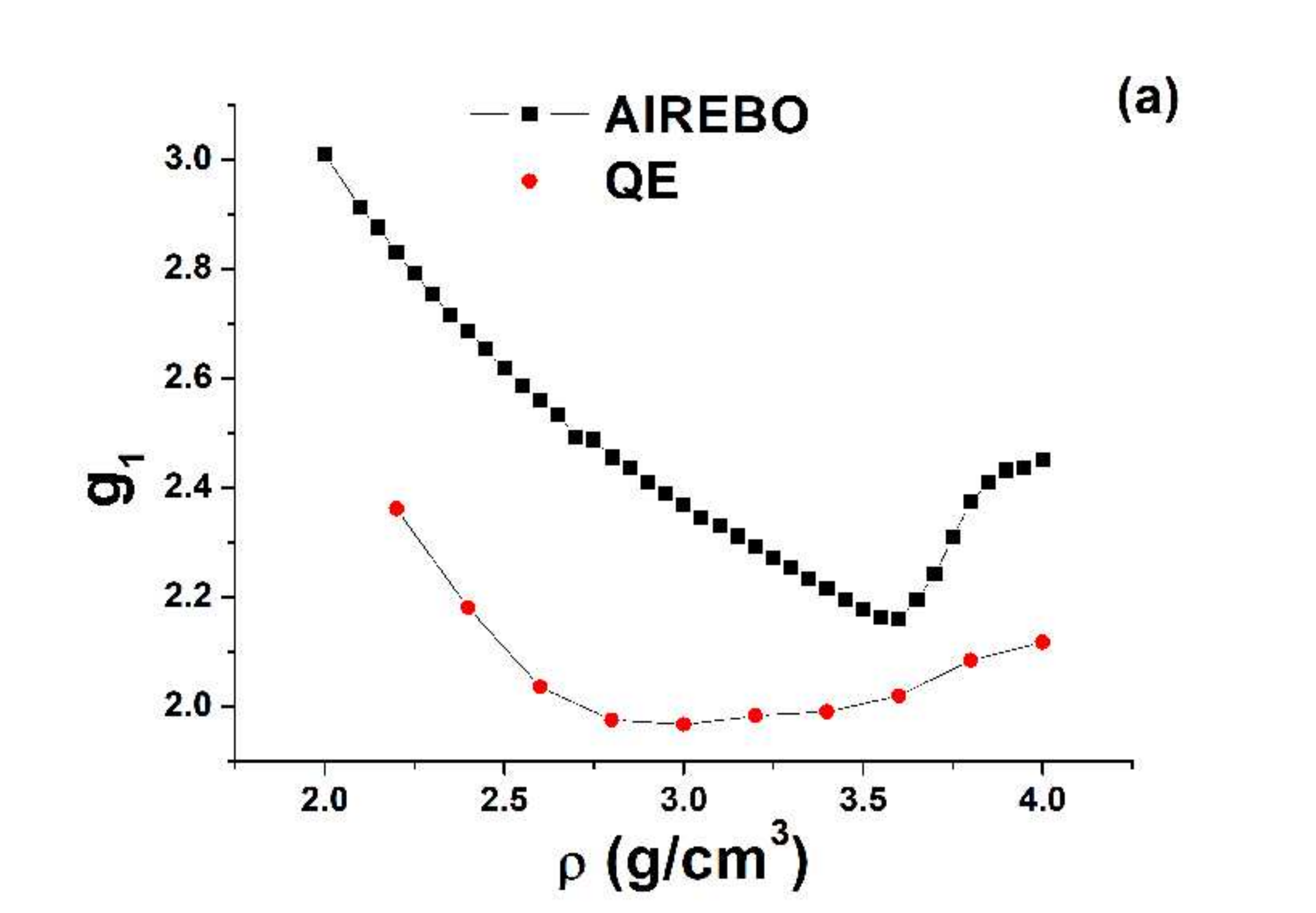}%

\includegraphics[width=6cm,height=6cm]{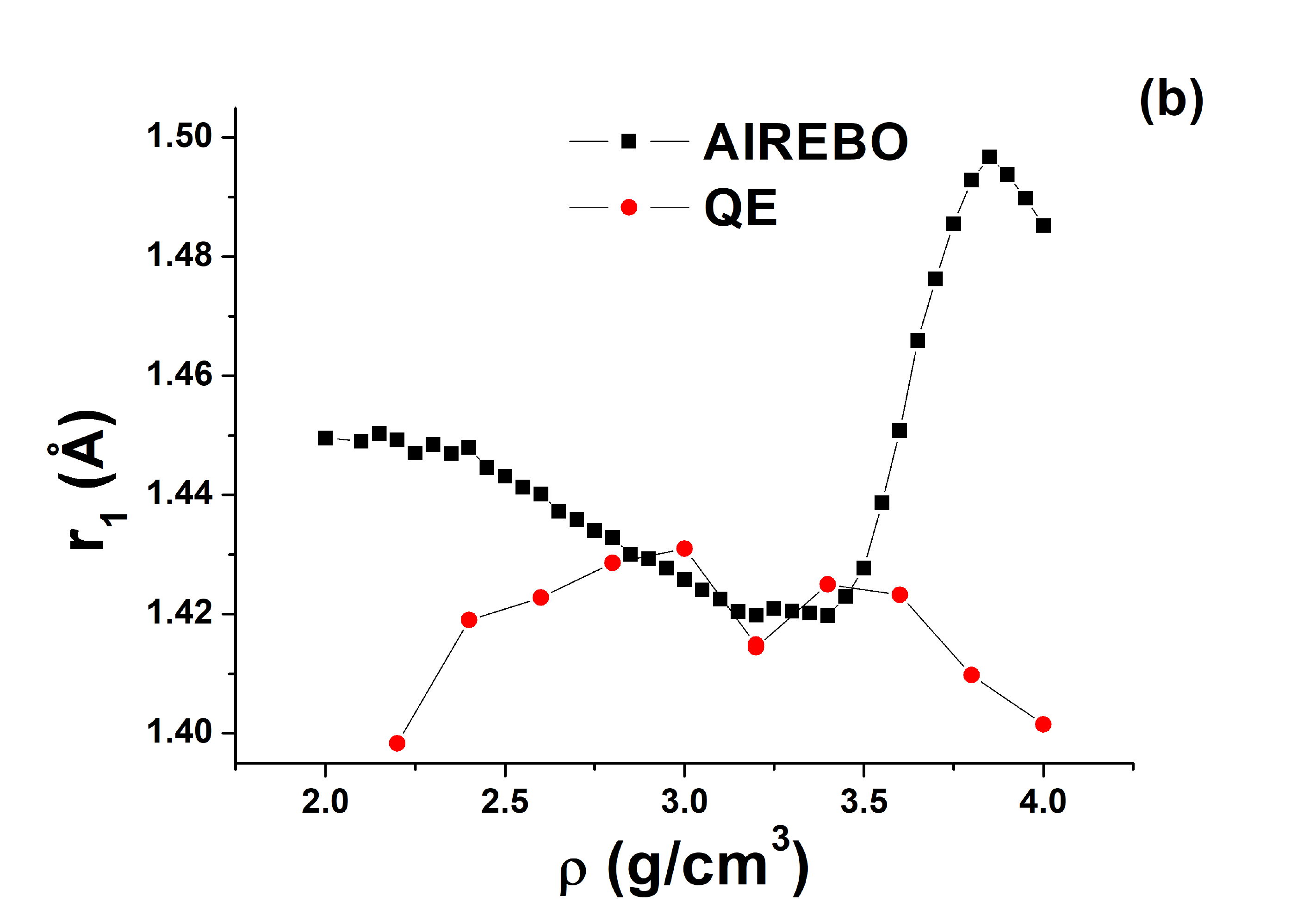}%

\caption{\label{g1r1}(a) Height of the first peak of the RDF of the AIREBO model at $T=6000$ K. (b) Position of the first peak of the RDF in the same system.}
\end{figure}

\begin{figure}
\includegraphics[width=6cm,height=6cm]{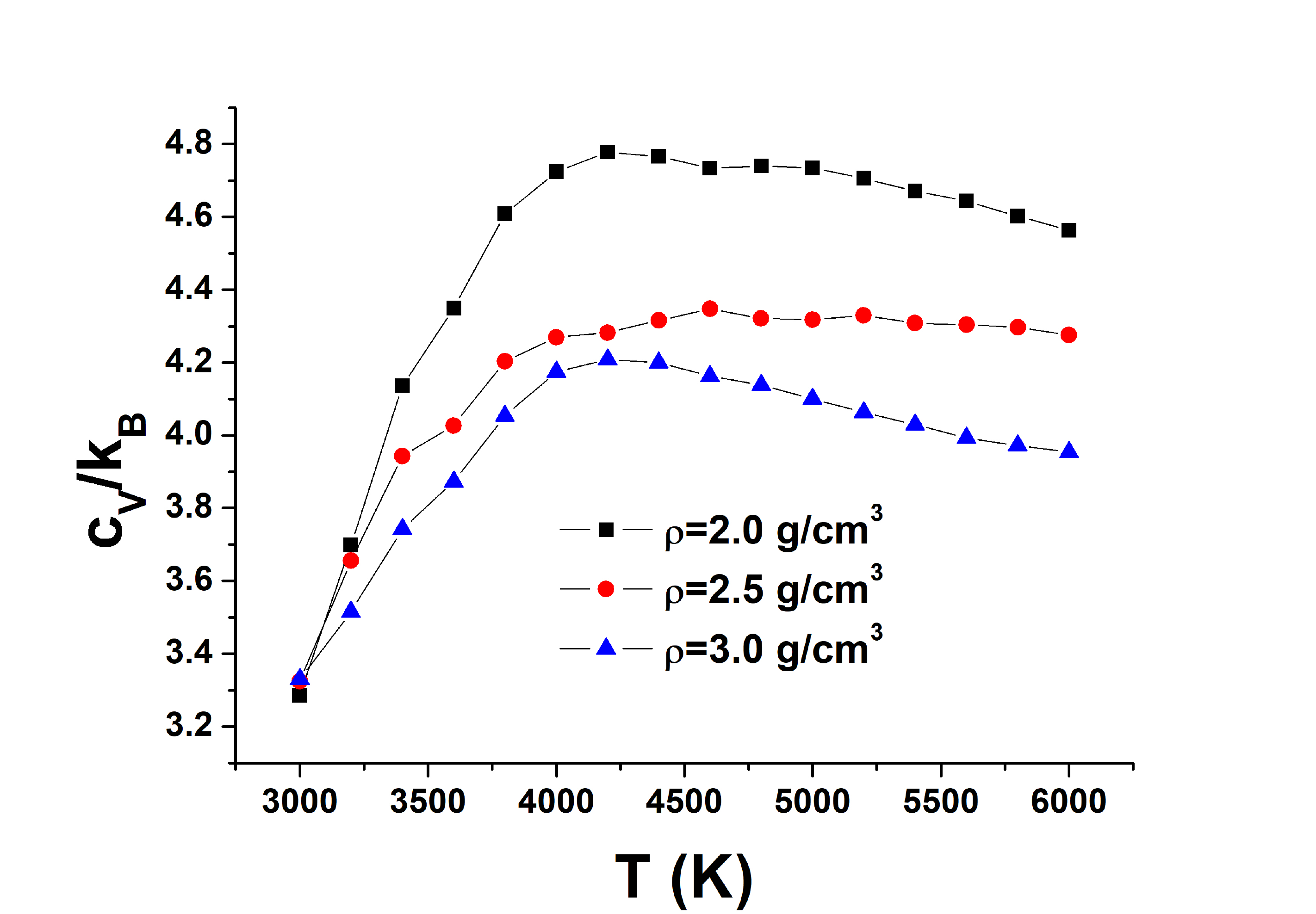}%

\caption{\label{cv} Specific heat at constant volume of the AIREBO model at three different densities.}
\end{figure}

A decrease in the position of the first minimum of the RDF $g_1$ means that a smooth structural crossover occurs in the system. One more attribute of the smooth crossover in the liquid phase can be seen in the system. The specific heat $c_V$ at constant volume of solids close to the melting line is $3k_B$ per particle (Dulong -- Petit law). The specific heat of liquids near the melting line is usually close to the same
value. It can be slightly larger due to the effects of anharmonicity, but it usually does not exceed $(3.3—3.4)k_B$ per particle. As shown in our recent papers \cite{cvlarge,cvlarge1}, a smooth structural crossover leads to a strong increase in the specific heat ($c_V
>3.5k_B$). Figure \ref{cv} shows the specific heat of AIREBO carbon along three different isochores. One can see that at all these densities, it exceeds $4k_B$ reaching a value as high as $4.78k_B$ per particle at
$\rho=2g/cm^3$ and $T=4200$ K. Summarizing an unusual behavior of
$g_1$ and high values of $c_V$, one can claim that a smooth structural crossover occurs in the system. This conclusion is interesting because most of the previous studies discussed the presence or absence of the LLPT in liquid carbon. The presence of the smooth crossover is a kind of an intermediate solution that allows
understanding the difficulty of interpretation of the data: some
properties are similar to those at a phase transition while others
are not.

We are not aware of any calculations of the melting line of diamond within the AIREBO model. However, comparing it to the results of LCBOPI+, one can say that the density $4g/cm^3$ should be close to the melting line of diamond or even in the metastable region below it. In this respect, it seems reasonable that liquid carbon approaching the diamond melting line changes its structure from graphite-like to diamond-like.
However, the RDFs of the system demonstrate at least two strange
features. First, the first minimum is as low as almost zero,
which is very atypical for liquids. Interestingly, the same
effect of the first minimum of the RDF was obtained in Ref. \cite{gap}
by using the Tersoff potential for carbon. Such deep minima of RDFs
are usually observed in an amorphous state (see, e.g., textbook \cite{ziman} for comparison of the RDFs of liquid and amorphous states. The experimental RDFs of amorphous carbon can be found in \cite{robertson}). In AIREBO carbon, such a deep minimum remains even at $T=6000$ K, which is about $1.65T_m$, i.e., far above the melting temperature. Second, there is an unusual kink at $r \approx 2 \AA$. Importantly, AIREBO includes so-called switching functions, which switch on the interaction if the distance between the atoms is within some cutoff distance and switch it off otherwise. The cutoff of the switching function of the AIREBO model is exactly 2 $\AA$ (Table I in Ref. \cite{airebo}), which allows us to suspect that this kink is artificially induced by the potential parametrization.

\begin{figure}
\includegraphics[width=8cm,height=8cm]{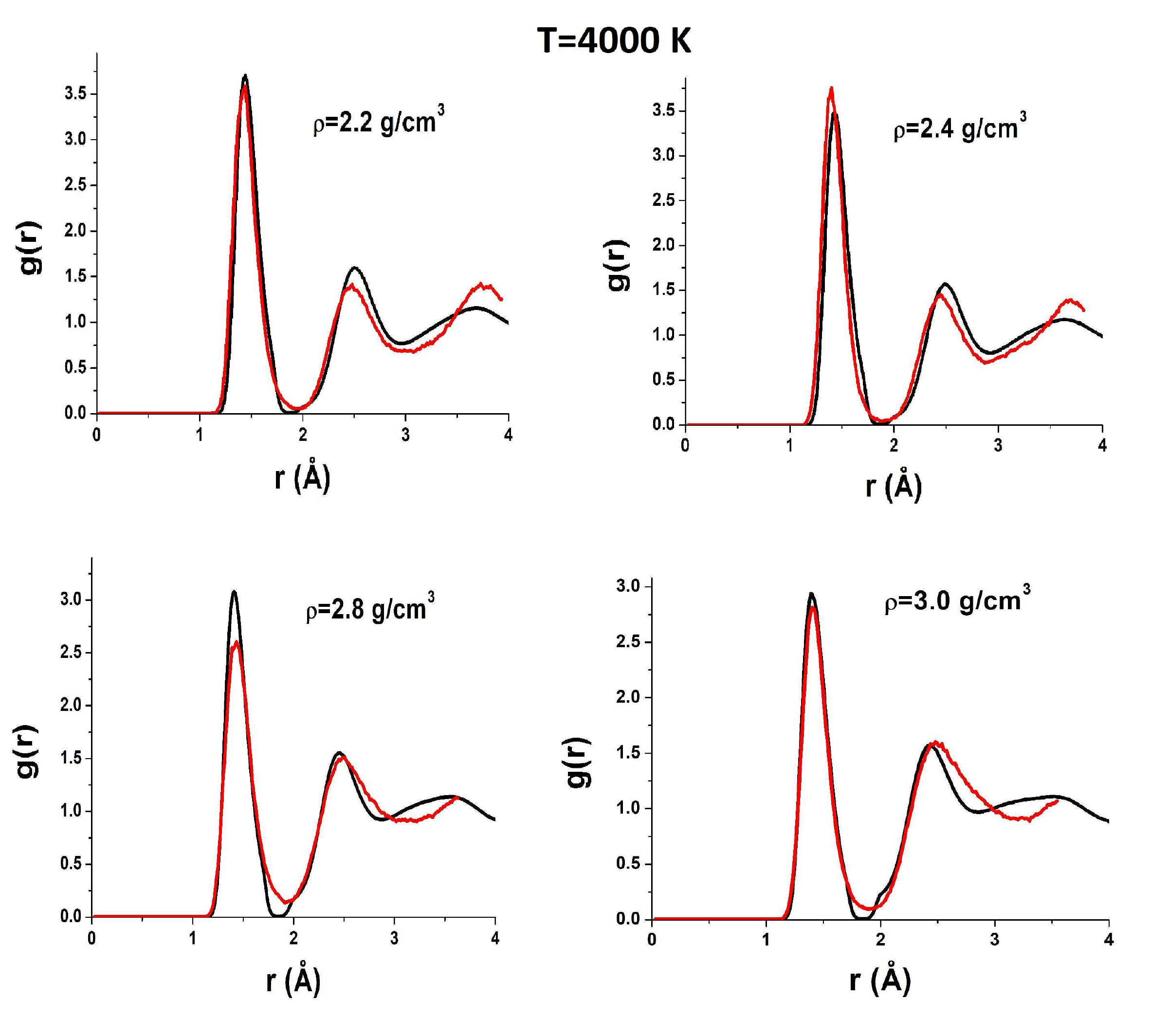}%

\caption{\label{rdf-t4000}Radial distribution functions of carbon obtained within the AIREBO model in comparison with those from ab initio molecular dynamics at $T=4000$ K and four different densities.}
\end{figure}

To reveal the origin of these effects of AIREBO RDFs, we compare them to those obtained in ab initio simulations. Figure \ref{rdf-t4000} shows
RDFs of liquid carbon obtained in the AIREBO model and ab initio simulation at $T=4000$ K, which corresponds to temperatures slightly above the melting line in the case AIREBO. One can see that the agreement between AIREBO and ab initio RDFs is relatively good at this temperature.
The main difference is in the second peak of the RDF. However, the kink of the AIREBO RDF at a high density starts to appear, while no kink is observed in ab initio RDFs. Importantly, ab initio RDFs at a low density
are also almost zero at the first minimum. However, as the density is increased, the first minimum of ab initio RDFs increases, while in the AIREBO model, it remains vanishingly small at all densities.

Figure \ref{rdf-t5000} shows AIREBO RDFs at $T=5000$ K in comparison with ab initio RDFs. One can see that both strange features are still present in AIREBO RDFs, but they are not observed in QE ones. In particular, the first minimum is finite at all densities.

\begin{figure}\label{rdf-t5000}
\includegraphics[width=8cm,height=8cm]{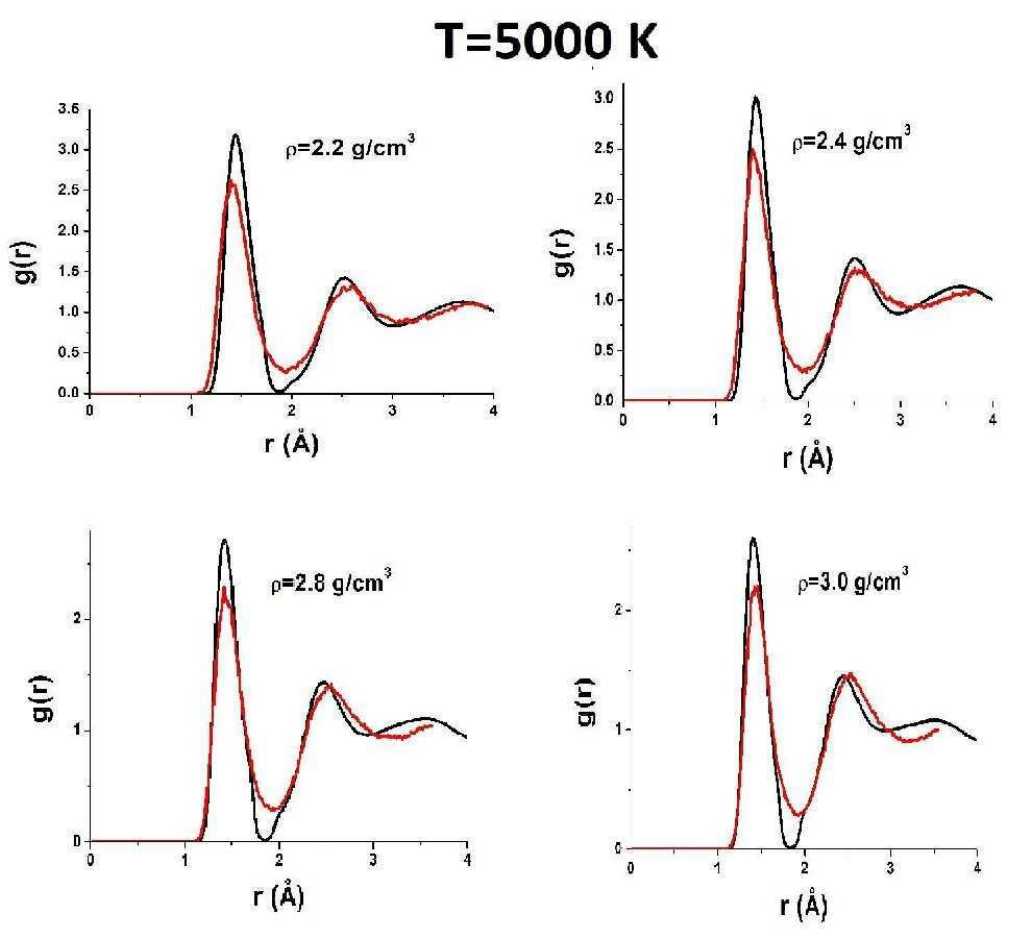}%

\caption{\label{rdf-t5000}Radial distribution functions of carbon obtained within the AIREBO model in comparison with those from ab initio molecular dynamics at $T=5000$ K and four different densities.}
\end{figure}

\begin{figure}
\includegraphics[width=8cm,height=8cm]{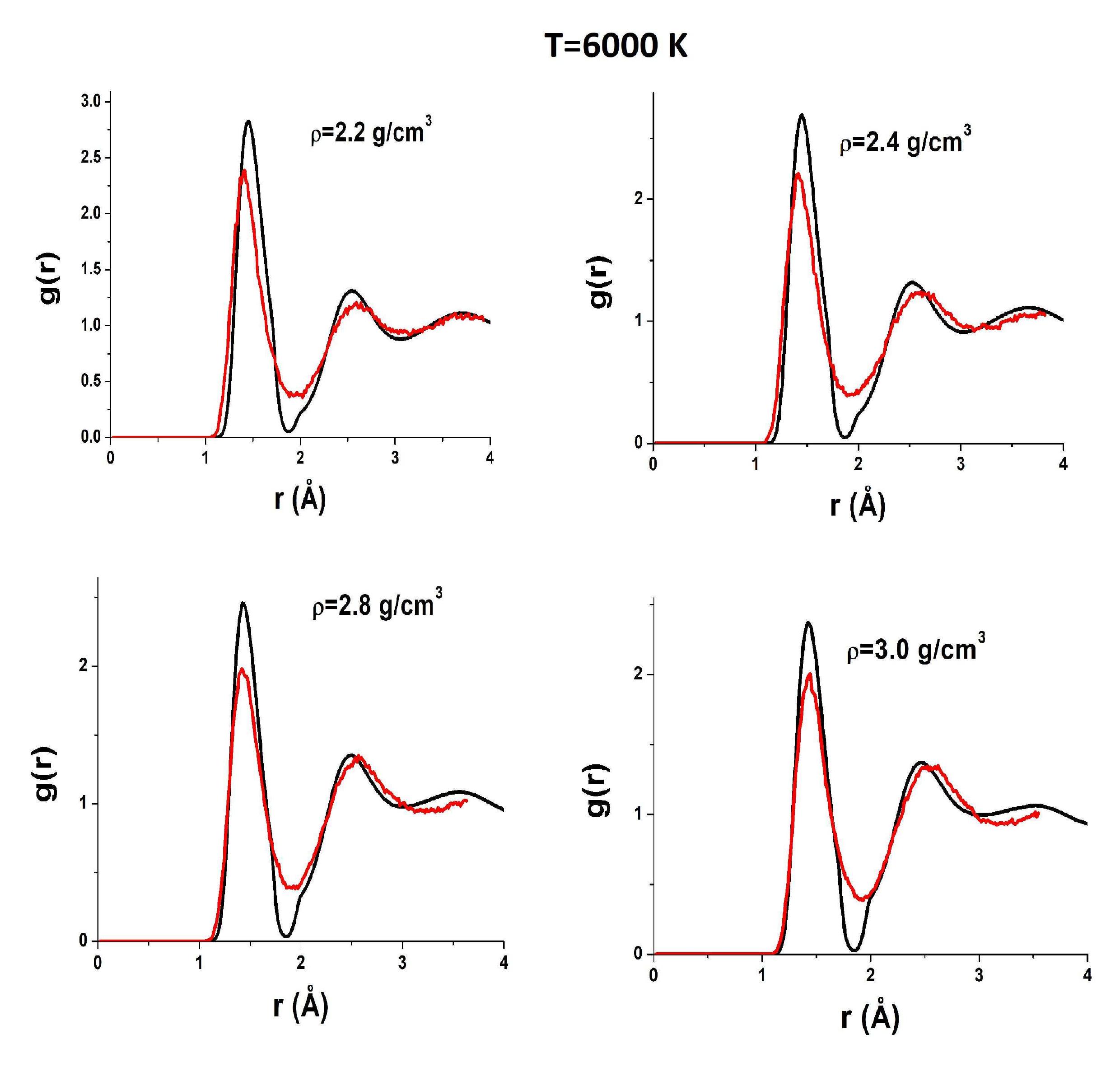}%

\caption{\label{rdf-t6000}Radial distribution functions of carbon obtained within the AIREBO model in comparison with those from ab initio molecular dynamics at $T=6000$ K and four different densities.}
\end{figure}

\begin{figure}
\includegraphics[width=8cm,height=8cm]{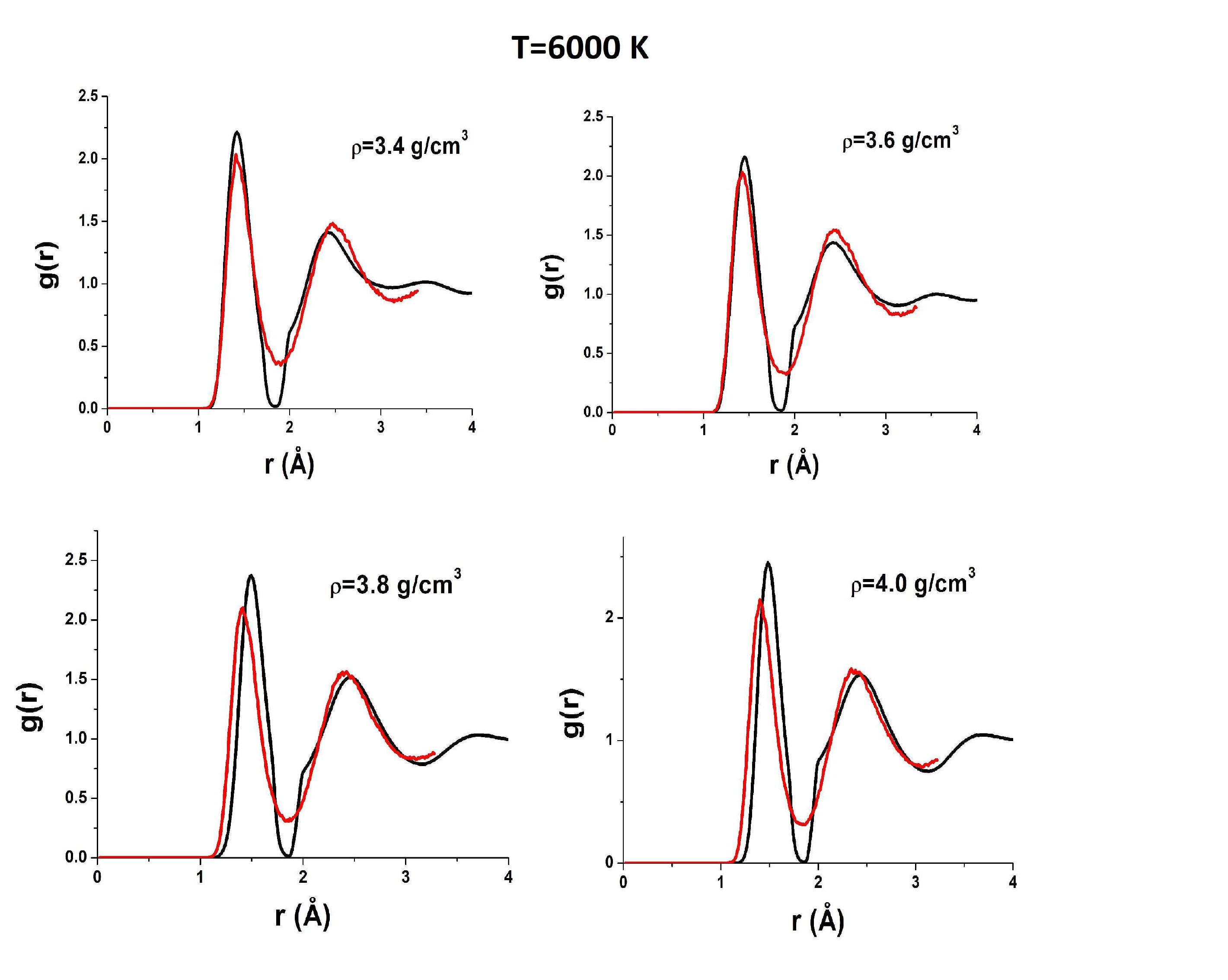}%

\caption{\label{rdf-t6000-hd}Radial distribution functions of carbon obtained withithe n AIREBO model in comparison with those from ab initio molecular dynamics at $T=6000$ K and high densities.}
\end{figure}

Figures \ref{rdf-t6000} and \ref{rdf-t6000-hd} show AIREBO RDFs in comparison with ab initio RDFs at low and high densities. It is seen again that the first minimum of AIREBO RDFs falls almost to
zero, while ab initio ones demonstrate finite values of the first
minimum at all densities. Moreover, one can see that the position of the first maximum of QE RDFs at high densities close to the melting line of diamond becomes smaller than that in the case of the AIREBO model. Therefore, the bond lengths in AIREBO and QE carbon become different at these densities. Figure \ref{g1r1} shows $r_1$ and $g_1$ obtained in quantum simulation in comparison with those within the AIREBO system. One
can see that the agreement is very poor. Anyway, it is seen that the behavior of ab initio RDFs is even more complicated than that of the AIREBO ones: the position of the first maximum $r_1$ increases with the density up to $\rho=3.2g/cm^3$ and its height decreases with an increase in the density up to $\rho=3.0g/cm^3$. Interestingly, while the difference between AIREBO and ab initio RDFs at $T=4000$ K is mostly at the second peak, the second peaks of AIREBO and ab initio $g(r)$ at $T=6000$ K are in good agreement, but the first peaks and especially the first minima are strongly different.

Many authors report the RDFs of liquid carbon obtained within different models. In Ref. \cite{gap}, a new potential called GAP is constructed by the machine learning methods. The authors of \cite{gap} calculate the RDFs with this GAP potential and with Tersoff one. In Ref. \cite{tb}, RDFs of carbon at $T=6000$ K are obtained by the tight-binding simulation technique. In Ref. \cite{colonna}, RDFs of LCBOPII carbon are calculated. The Tersoff and AIREBO empirical potentials are constructed using the low temperature data for graphite, diamond and hydrocarbons. The LCBOPII and GAP models are fitted to ab initio data. The tight-binding method is in some sense similar to ab initio simulations. Comparing the RDFs of different models, one can see that the Tersoff and AIREBO potentials give an unphysically low first minimum while the ab initio, tight binding and empirical potentials fitted to the ab initio data give liquid-like RDFs of liquid carbon. For this reason, we can conclude that models based on the low temperature data such as Tersoff and AIREBO do not describe liquid carbon properly. It also leads to the conclusion that these models do not describe the melting line of graphite, because to precisely describe a phase transition, the models should precisely describe both phases. For this reason, we believe that the LLPT observed in AIREBO model is an artificial effect of the model. However, the ab initio simulations still confirm the presence of a smooth structural crossover in liquid carbon, which can be seen from the behavior of the coordination number, $g_1$ and $r_1$. At the same time, as shown in Ref. \cite{dozhdikov}, the accuracy of ab initio calculations is also doubtful. For this reason, the presence or absence of this structural crossover requires further justifications.

\section{4. Sublimation ("melting") of graphene within the AIREBO model}

Finally we discuss melting of graphene modeled with the AIREBO potential.

It is important to note that there is a difference in simulation methodology of  melting of graphite and graphene. The most principal item is that melting of graphite is always simulated at a high enough external pressure, which is above the pressure at the triple point ($P_{tr}=0.016$ GPa). For instance, the lowest pressure in Ref. \cite{orekhov} is 2 GPa. In the case of graphene, the simulations are performed at a nearly zero pressure. In both cases of LCBOPII and AIREBO, a graphene sample surrounded by vacuum is simulated at a pressure fluctuating around zero. The initial temperature is 300 K and the temperature increases at a constant rate during the simulation. The melting temperature is detected by structural changes in the system in Refs. \cite{zakh} and \cite{orgr}. Extrapolation to lower temperatures is made in Ref. \cite{los}.

It is known that no melting is possible if the pressure is below the triple point. In the case of almost zero pressure, the substance should be sublimated, i.e., be transformed from crystal directly to the gas phase. Moreover, the sublimation temperature at $P=0$ Pa is also zero. In computer simulations, the pressure is never exactly equal to the desired value. Indeed, strong fluctuations always occur in simulations. For this reason, one can expect that a low positive pressure is indeed simulated in all cited publications. In our simulations, fluctuations of the
pressure are within $10^{-3}$ GPa.  However, one can still expect that gaseous carbon appears in such simulations instead of liquid. Although the experimental studies of gaseous carbon are extremely difficult due to very high temperatures, several works report that gaseous carbon consists
of clusters of several atoms. In particular, in \cite{leider}, the partial pressure of clusters of different sizes is estimated. It is shown that the major contribution comes from the clusters of three carbon atoms. In this respect, it seems reasonable to assume that the transition from graphene to gaseous carbon (i.e., sublimation) is indeed observed in simulations.

\begin{figure}
\includegraphics[width=8cm,height=8cm]{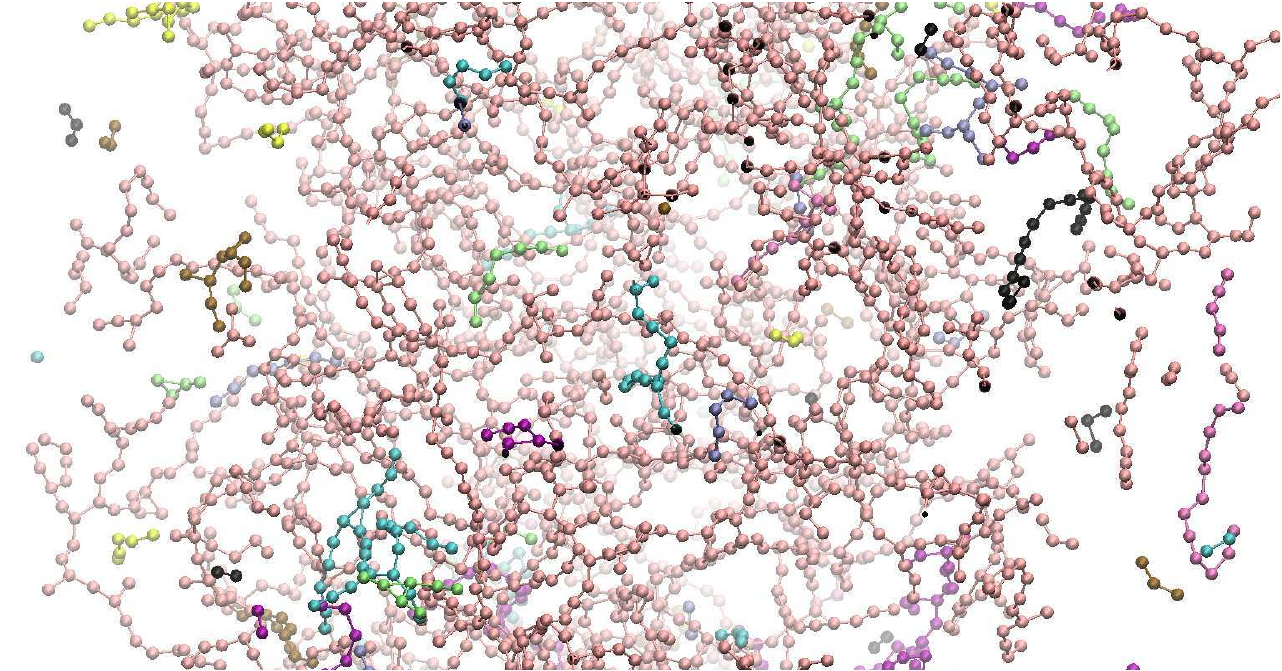}%

\caption{\label{gr8000} Snapshot of the system after "melting". The initial structure is graphite. The temperature is $T=8000$ K. The system in divided into clusters (see the text). Different clusters are marked by different colors. Note that since the number of clusters is large,
different clusters can be drawn by the same color.}
\end{figure}

In our work, we simulate the thermal decomposition of both graphite and graphene at zero pressure. In both cases, we simulate the system at 8000 K starting with graphene or graphite as initial structures. In the case of graphite, one should observe sublimation, since zero pressure is apparently below the triple point pressure. The aim of this study is to
compare the structures of the system after the collapse of the crystal structure in the cases of graphene and graphite. The time step is 0.2 fs. The graphene and graphite samples consist of 6400 and 4000 atoms, respectively. The initial lattice constants are taken as those under ambient conditions. Figure \ref{gr8000} (a) shows a snapshot of a graphene sample after "melting". In agreement with Refs.  \cite{zakh,los,orgr}, the system consists of linear chains. The first minimum of the RDF is $r_c=1.95$ $\AA$. We divide the system into clusters. We assume that two particles belong to the same cluster if
they are separated by a distance less than $r_c$. Figure \ref{gr8000} (b) shows the probability distribution of the cluster sizes $P(N_c)$ ($N_c$ is the size of the cluster). One can see that most of the clusters consist of three particles in agreement with expectations for gaseous carbon. The largest cluster consists of 56 atoms.

\begin{figure}
\includegraphics[width=8cm,height=8cm]{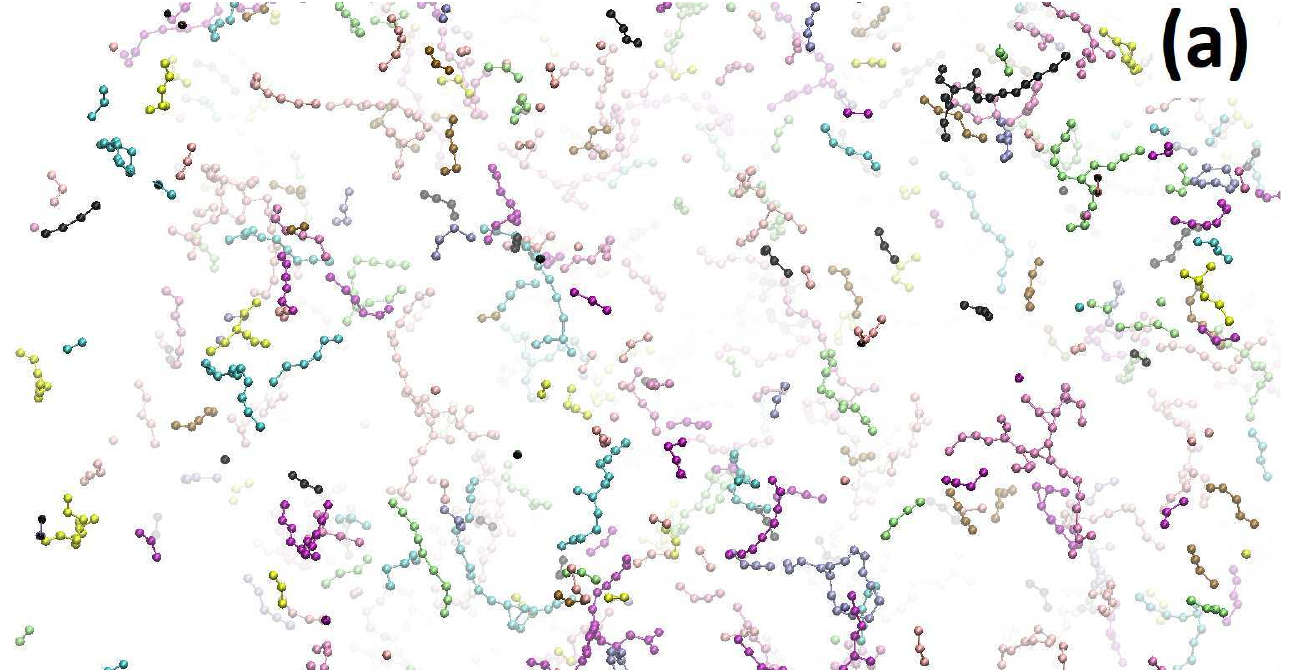}%

\includegraphics[width=8cm,height=8cm]{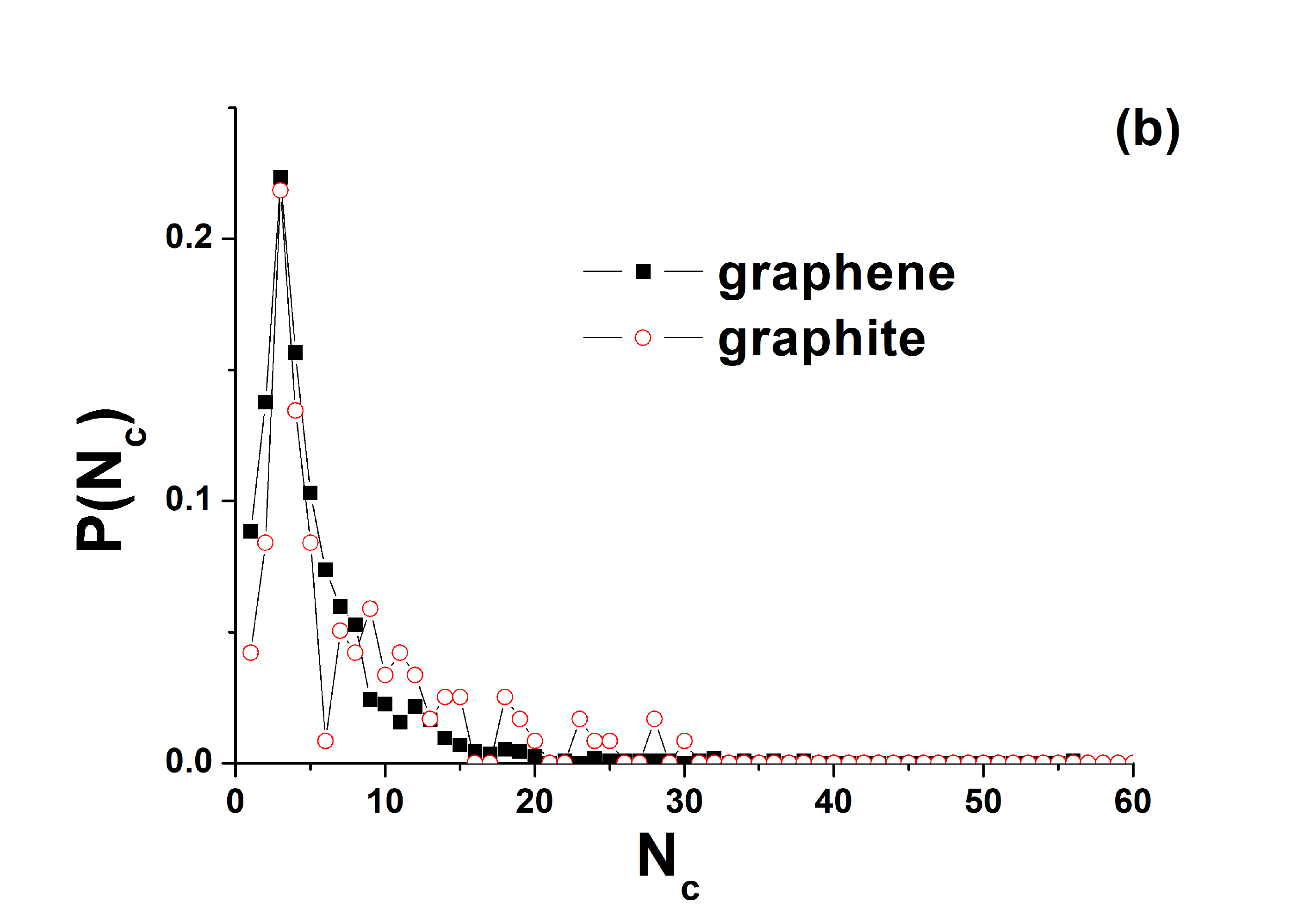}%

\caption{\label{graphite8000} (a) Snapshot of the system after "melting". The initial structure is graphene. The temperature is $T=8000$ K. The system in divided into clusters (see the text). Different clusters are marked by different colors. Note that since the number of clusters is large, different clusters can be drawn by the same color. (b) Probability
distribution of the cluster sizes in two systems with initial configuration of graphene and graphite.}
\end{figure}

We also perform the simulation of graphite at $P \approx 0$ GPa and $T=8000$ K and find that the sample is also transformed into the disordered state (Fig. \ref{gr8000}). The first minimum of the RDF is also at $r_c=1.95$. Dividing the system into clusters, we find that the majority of the atoms belong to a single cluster (3020 particles). However, other particles form linear chains. The size distribution of these chains is given in Fig. \ref{graphite8000} (b) in comparison with that for the system started with graphene. One can see that the distributions for the two cases are very similar. In particular, most of the chains in both cases consist of three particles.

It is very important that the volume of the system in both cases (starting with graphene or graphite) continues to increase with time even after 150-ns-long simulation. Therefore, the system is still out of equilibrium and the structure will continue to change if we simulate the system for a longer time. This behavior allows us to conclude that both graphite and graphene being heated at zero pressure are transformed into gaseous carbon, but the equilibration of the system is extremely long and, for this reason, no firm conclusions can be made even after 150-ns-long simulation. One can see that the equilibration of the system is extremely slow. This can explain why the system is "melted" at higher temperatures than usual graphite at high pressures. Experimentally, the sublimation of graphite starts at about 3000 K \cite{sheindlin}. Extremely strong overheating of graphene in simulation should also originate from the limitation of the simulation time.

The experimental estimates of the gas--liquid--graphite triple point are from about 3000 to 5000 K and $P=0.016$ GPa \cite{savv}. We emphasize that a single sheet of carbon atoms in the simulation of "melting" of graphene is placed in a box of vacuum; i.e., the whole system is indeed three-dimensional. Moreover, the decomposition of the graphene sheet proceeds via the formation of chains of atoms that are not in the graphene sheet. The final fluid phase is also a three-dimensional object. In this respect, one cannot consider graphene as a two-dimensional system, since out-of-plane effects are of fundamental importance. It would be more correctly to represent graphene as a limiting case of graphite consisting of a single layer. In principle, the triple point of such a single layer can be shifted with respect to that of bulk graphite. However, there is no reason to expect that it should be strongly different from the latter. As pointed out in Ref. \cite{orekhov}, the melting of graphite is related to bonding of atoms with atoms of other planes. No such mechanism is possible in the case of graphene. In this case, melting should be attributed to the breaking of the bonds. However, one should not expect that the temperature of bond breaking is too much higher than the melting point of graphite. For example, experimental observations show that graphite starts to actively sublimate already at $T=3000$ K \cite{sheindlin,savv}.

As far as we know, the triple point has been calculated within neither the AIREBO model nor the LCBOPII model. However, since the melting temperatures of these systems do not depend on the pressure, one should conclude that the temperature of the triple point should be the same as the melting temperature of graphite ($T_t=3640$ K for AIREBO and $T_t=4250$ K for LCBOPII). The "melting" (sublimation) points of graphene are $T_m=4900$ K for AIREBO and $T_m=4510$ K for LCBOPII. These sublimation temperatures are above the temperatures of the triple point of graphite, which does not make sense.


\section{5. Simulation of the melting of graphene under hydrostatic compression}

In order to make the simulation of graphene melting more realistic, we employ another scheme of simulation, which basically repeats the experimental setup. We simulate layers of graphene immersed in the argon atmosphere. Importantly, real experiments always contain some medium around the sample. The medium should be chemically inert and weakly interact with the atoms of the sample. Argon satisfies all these requirements and can serve as a good medium maintaining the desired pressure. This atmosphere allows us to support the pressure in the system at the desired level. We simulate the system at $P=0.1$ GPa, which
is well above zero pressure used in previous simulation works. Moreover, this pressure is several times higher than the pressure of the triple point of graphite. The melting temperature of graphite shows only a very weak pressure dependence. The pressure dependence of the melting line of graphene is unknown. However, the derivative $\partial T /\partial P$ of all inorganic substances does not exceed 30 $K/kbar$. For this reason, one should expect that the melting temperature at $P=0.1$ GPa does not exceed the melting temperature at lower pressures by more than 10--20 K.
Importantly, two-dimensional systems demonstrate the same features of the phase diagram as crystal, liquid and gas phases, which merge at the triple point. The triple point of graphene is unknown. However, one may expect that it is close to that of graphite. For this reason, one may expect that the pressure $0.1$ GPa is above the triple-point pressure of graphene as well. We simulate graphene in the argon atmosphere in an isobaric--isothermal ensemble at different temperatures ranging from $T=300$ K to $T=4000$ K with the step $\Delta T =100$ K and monitor when the crystalline structure begins to break. The system is simulated for 50 ns. Although some effects of finite simulation time can take place in the system, we believe that this simulation period should be enough to roughly determine the melting point. A graphene sheet consisting of
5400 carbon atoms was surrounded by 7000 argon atoms. The argon interaction is modeled by a Lennard-Jones potential with the parameters
$\varepsilon_{Ar} =0.01034$ eV and $\sigma_{Ar}=3.4$ $\AA$. The argon--carbon interaction is obtained by the Lorentz--Berthelot mixing rule $\varepsilon_{Ar-C}=(\varepsilon_{Ar}\varepsilon_C)^{0.5}$ and $\sigma_{Ar-C}=0.5(\sigma_{Ar}+\sigma_C)$,
where the parameters for carbon are taken as the Lennard-Jones parameters of the AIREBO potential \cite{airebo} $\varepsilon_C=0.00284$ eV and $\sigma_{C}=3.4$ $\AA$. This gives $\varepsilon_{Ar--C} =0.00542$ eV and $\sigma_{Ar--C}=3.4$ $\AA$.

Our simulations show that the crystalline structure breaks at $T=3700$ K. This temperature is consistent with the melting temperature of graphite from \cite{orekhov}, but is much lower than $4900$ K where sublimation was observed in the AIREBO model of graphene \cite{orgr} and $T=4510$ K in the simulation of LCBOPII graphene \cite{zakh,los}. Most probably, the simulation setup based on simulation of the graphene sheet in vacuum is incorrect, which leads to huge overestimation of the melting temperature.

\begin{figure}
\includegraphics[width=8cm,height=8cm]{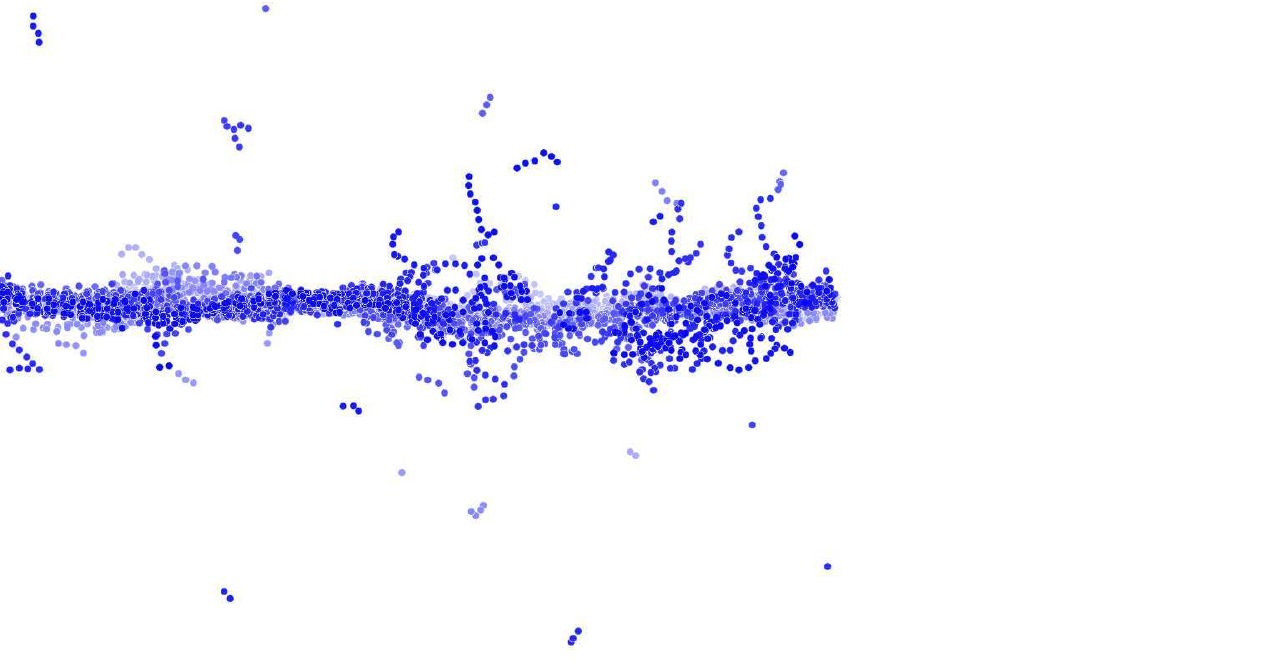}%

\includegraphics[width=8cm,height=8cm]{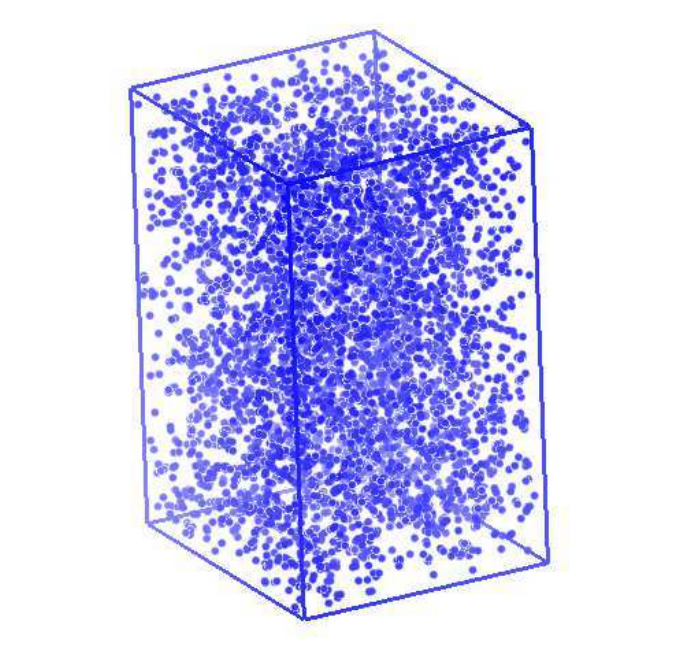}%

\caption{\label{gr3700} (a) Snapshot of the carbon part of the system with argon at initial stages of melting. (b) Snapshot of carbons in the system when melting is completed.}
\end{figure}

Figure \ref{gr3700} (a) shows a snapshot of carbons in the system after initial stages of the collapse of the crystalline structure. Figure \ref{gr3700} (b) shows a snapshot of the system when melting is complete. One can see that melting at the initial stage proceeds via
the formation of linear chains, which is similar to the case of graphene sublimation. At the final stage, the system looks uniform, as should be in liquid.

\section{6. Conclusions}


To summarize, we have reported a new simulation study of the melting behavior of graphite and evaporation and melting of graphene. Our data have been critically compared to the results of previous studies. Analysis of our results and comparison with the literature sources have provided the following conclusions.

i) The empirical potentials for carbon can be divided into two groups. Some of them (AIREBO, Tersoff potential, Brenner potential) are fitted to experimental data on carbon and hydrocarbons at low temperatures. These models fail to reproduce the structure of liquid carbon above the melting line. For this reason, they also do not describe the melting line
of graphite properly.

ii) The AIREBO potential gives an LLPT at high densities where the graphite-like liquid is transformed into the diamond-like one. However, this LLPT is not observed in ab initio simulations; it is most obviously a property of the model rather than of real carbon.

iii) Both AIREBO and ab initio simulations give an unusual dependence of the position and height of the first maximum of the RDF of liquid carbon. The same effect is observed in the tight-binding simulation of liquid carbon (Ref. \cite{tb}). This circumstance means that a smooth structural crossover occurs in liquid carbon. This smooth LLPT is also observed in ab initio simulations. Importantly, the presence of this smooth LLPT means that the melting line should demonstrate a diffuse maximum, which
is in agreement with the classical results obtained by Bundy \cite{bundy}.

iv) The AIREBO and LCBOPII models predict that the sublimation ("melting") temperature of graphene is much higher than the melting temperature of graphite. Although this effect is observed in simulations, it seems to be a consequence of an incomplete equilibration of the system. Indeed, the simulations of "melting" of graphene are performed at the pressure where evaporation should be observed instead of melting.
The structure of the phase observed in simulations of "melting" of graphene seems to be consistent with predictions for gaseous carbon. Under the assumption that the "melting" temperature of graphene is higher than that for graphite, we arrive at a paradoxical conclusion that the sublimation temperature of graphene is higher than the melting temperature of graphite.


v) A new simulation scheme has been proposed to study the genuine melting of graphene: immersing the graphene sheet into the argon atmosphere with a pressure above the triple-point pressure of graphite. This scheme gives a much lower melting temperature than the sublimation temperature of graphene in vacuum. The melting temperature obtained in our simulation is in agreement with the melting temperature of graphite obtained with the same model of carbon.



\bigskip

This work was carried out using computing resources of the federal
collective usage center "Complex for simulation and data
processing for mega-science facilities" at the NRC "Kurchatov
Institute", http://ckp.nrcki.ru, and supercomputers at the Joint
Supercomputer Center of the Russian Academy of Sciences (JSCC
RAS). The work was supported by the Russian Science Foundation (Grants 19-12-00111 and 19-12-00092 for investigations of graphite and graphene, respectively).

\end{document}